\documentclass[pre,onecolumn,reprint,aps,superscriptaddress,longbibliography]{revtex4-1}
\usepackage{graphicx}%
\usepackage{dcolumn}%
\usepackage{bm}%
\usepackage{color}
\usepackage{multirow}
\usepackage[normalem]{ulem}
\usepackage{amsmath}
\usepackage{enumerate}
\usepackage{amsfonts}
\usepackage{epsfig}
\usepackage{graphicx}
\usepackage{cancel}
\usepackage[caption=false]{subfig}
\usepackage{floatrow}
\usepackage{subfig}
\usepackage{bbm}
\newcommand{\be}{\begin{equation}}
\newcommand{\ee}{\end{equation}}
\newcommand{\bea}{\begin{eqnarray}}
\newcommand{\eea}{\end{eqnarray}}

\definecolor{green}{rgb}{0.0, 0.44, 0.0}
\definecolor{red}{rgb}{1.0, 0.13, 0.32}
\definecolor{blue}{rgb}{0.06, 0.2, 0.65}
\definecolor{darkgreen}{rgb}{0,0.5,0}
\definecolor{darkblue}{rgb}{0,0,0.6}
\definecolor{purple}{rgb}{0.4,.2,0.7}
\definecolor{magenta}{rgb}{1.0,0.0,1.0}
\usepackage[colorlinks=true,citecolor=darkgreen,linkcolor=purple,urlcolor=purple]{hyperref}

\def\le{\left}
\def\ri{\right}

\def\T{\mathcal{X}}
\def\gI{g^{\mathrm{I}}}
\def\gII{g^{\mathrm{II}}}
\def\dS{d_{0}}
\def\dH{d_{\mathrm{H}}}
\def\dC{d_{\mathrm{cas}}}
\def\du{d_{\mathrm{u}}}
\def\dcol{d_{\mathrm{col}}}
\def\lamAH{\lambda_{\mathrm{aH}}}

\begin{document}

\author{Patrick Charbonneau} 
\affiliation{Department of Chemistry, Duke University, Durham,
North Carolina 27708, USA}
\affiliation{Department of Physics, Duke University, Durham,
North Carolina 27708, USA}

\author{Yi Hu}
\affiliation{Department of Chemistry, Duke University, Durham, 
North Carolina 27708, USA}

\author{Archishman Raju}
\affiliation{Laboratory of Atomic and Solid State Physics, Cornell University, Ithaca,
New York 14853, USA}

\author{James P.~Sethna}
\affiliation{Laboratory of Atomic and Solid State Physics, Cornell University, Ithaca,
New York 14853, USA}

\author{Sho Yaida}
\thanks{\href{mailto:shoyaida@fb.com}{shoyaida@fb.com} (effective as of August 13, 2018).}
%\email{shoyaida@fb.com}
\affiliation{Department of Chemistry, Duke University, Durham, 
North Carolina 27708, USA}

\title{Morphology of renormalization-group flow for the de Almeida-Thouless-Gardner universality class} 

\begin{abstract}
A replica-symmetry-breaking phase transition is predicted in a host of disordered media. The criticality of the transition has, however, long been questioned below its upper critical dimension, six, due to the absence of a critical fixed point in the renormalization-group flows at one-loop order. A recent two-loop analysis revealed a possible strong-coupling fixed point but, given the uncontrolled nature of perturbative analysis in the strong-coupling regime, debate persists. Here we examine the nature of the transition as a function of spatial dimension and show that the strong-coupling fixed point can go through a Hopf bifurcation, resulting in a critical limit cycle and a concomitant discrete scale invariance. We further investigate a different renormalization scheme and argue that the basin of attraction of the strong-coupling fixed point/limit cycle may thus stay finite for all dimensions.
\end{abstract}

\maketitle

\section{Introduction}\label{intro}
Quenched disorder often leaves conspicuous marks on a system's macroscopic behavior. For instance, quenched impurities can localize excited states~\cite{Anderson58,BAA06,OH07,PH10}
%Lifshitz tails~\cite{Lifshitz64};
and consequently turn metals into insulators with anomalous transport properties~\cite{Mott69,AHL71,ES75}. When coupled to an order parameter, extrinsic disorder can also destroy the would-be long-range order, altering its lower critical dimension~\cite{IM75}. Counterintuitively, disorder can also give rise to long-range order, albeit in the subtle, amorphous manner that breaks the permutation symmetry among fictitious replicas~\cite{Parisi79}. Although initially considered a fairly exotic proposal, this replica symmetry breaking (RSB) phenomenon has since found core applications in different fields of science~\cite{MPV87}.

The nature of the RSB phase transition, however, remains controversial. While its existence and criticality are unquestionable in a wide range of infinite-dimensional mean-field models ranging from spin to structural glasses~\cite{EA75,SK78,AT78,Parisi79,MPV87,Guerra03,Talagrand06,Gardner85,GKS85,CKPUZ14}, some have suggested that the RSB phase is completely washed out in any finite-dimensional, short-ranged models~\cite{NS03}.
Especially the droplet/scaling scenario~\cite{McMillan84,BM85,FH86,HF87,BM87,FH88} proposes that there cannot be infinitely many incongruent pure states in realistic finite-dimensional models (see, however, Ref.~\cite{WF06}).
Others posit that the transition survives down to the upper critical dimension, $\du=6$, but disappears below it~\cite{MB11}. This second proposal, in particular, is rooted in the absence of a critical fixed point in the renormalization-group (RG) flow equation at one-loop level below $\du$~\cite{BR80,PTD02,BU15}. The discovery of the Gardner transition in structural glass formers~\cite{CKPUZ14} has rekindled interest in this debate~\cite{BU15,AB15,RUYZ15,BU16,AB17,JY17}, and a recent two-loop RG analysis~\cite{CY17} challenges above proposals by identifying a strong-coupling critical fixed point that is invisible at one-loop order, just as is the case for a class of non-Abelian gauge theories~\cite{Caswell74,BZ82}. While the validity of the two-loop analysis in the strong-coupling regime can be questioned, it nonetheless provides a potentially viable description of the critical RSB transition in three-dimensional systems.

The difficulty associated with capturing the fate of strong-coupling fixed points through perturbative methods is well known. Even for the Ising universality class, the minimal RG equation without resummation results in the Wilson-Fisher fixed point for $d=2$ and $3$ being present at one-loop, three-loop, and five-loop orders, but absent at two-loop and four-loop orders. Only after applying a certain class of resummation schemes does the existence of the fixed point become independent of loop order~\cite{BNGM76}. In the Ising case, the pre-existing experimental and theoretical evidences of criticality in two and three dimensions, together with the striking agreement of the one-loop exponents with those in three dimensions, made it clear that the qualitative change in the unresummed results was not a fundamental concern. For the critical RSB phase transition as well, a similar aggregation of evidences from theories, experiments~\cite{Weissman93,Mydosh14,SD16,GLL18}, and simulations~\cite{JANUS12,JANUS14,BCJPSZ16,SBZ17,CCFTvdN18,SZ18} will be needed to reach a steady state of understanding for the true fixed point structure. In attaining such an understanding, it is especially instructive to examine the nature of the transition as a function of spatial dimension, $d$, as was instrumental for studying the Ising universality class, percolation~\cite{SA14,BH00}, the glass problem~\cite{RN7524,RN12133}, and many others~\cite{GoldenfeldBook}. Here, we thus closely analyze higher-loop RG flow equations in varying dimensions.

It is important to emphasize that the intent of the paper is not to provide a conclusive answer to the nature of the fixed point structure in finite dimensions. That answer will most likely require a concerted and sustained effort in developing various theoretical machineries such as higher-loop calculations~\cite{Gracey15} with sophisticated resummation schemes~\cite{ZinnJustin10,DU15}, nonperturbative RG~\cite{Polchinski84,Wetterich93,Morris94,BTW02,TT04,Delamotte12}, and conformal bootstrap~\cite{RRTV08,EPPRSV12,Gliozzi13,EPPRSV14}, as well as experiments and simulations. Instead, our intent here is to suggest a few viable physical scenarios that have heretofore been missed within the confine of the one-loop analysis.

The organization of the paper is as follows. In Sec.~\ref{minimal}, we analyze the minimal two-loop RG flow equation and in particular find that, as $d$ varies, the fixed point goes through a Hopf bifurcation, resulting in a limit cycle with discrete scale invariance. There, a controversy in $d>\du$~\cite{MR18} is also addressed. We then employ a coordinate-transformed RG scheme in Sec.~\ref{normal}, within which the basin of attraction of the critical fixed point/limit cycle stays finite for all $d$, in contrast to previously reported scenarios~\cite{CY17,MR18}. We then briefly conclude in Sec.~\ref{conclusion}.

\section{Minimal two-loop RG}
\label{minimal}
\begin{figure}[t]
\centerline{
\subfloat[One-loop]{\includegraphics[width=0.48\textwidth]{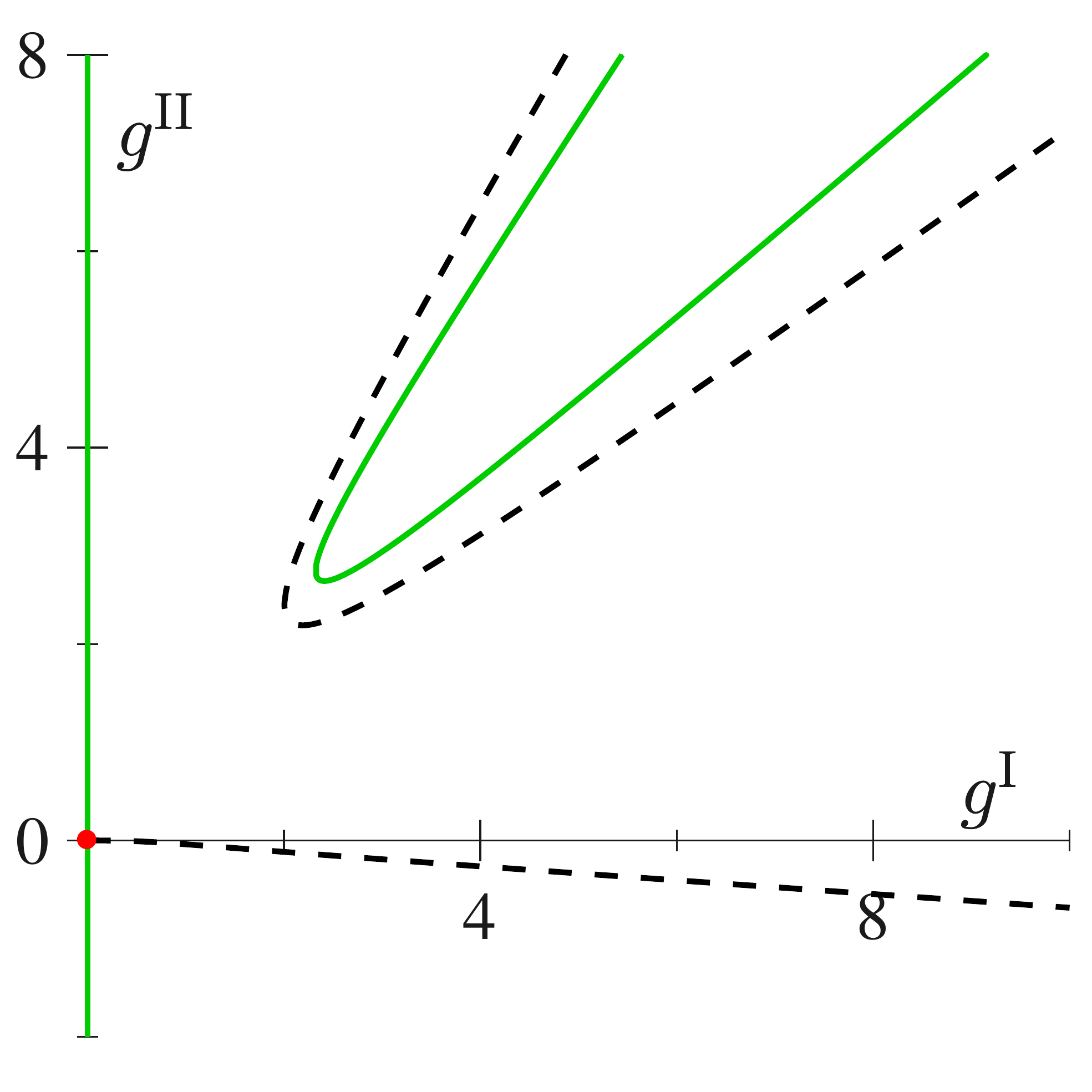}}\quad
\subfloat[Two-loop]{\includegraphics[width=0.48\textwidth]{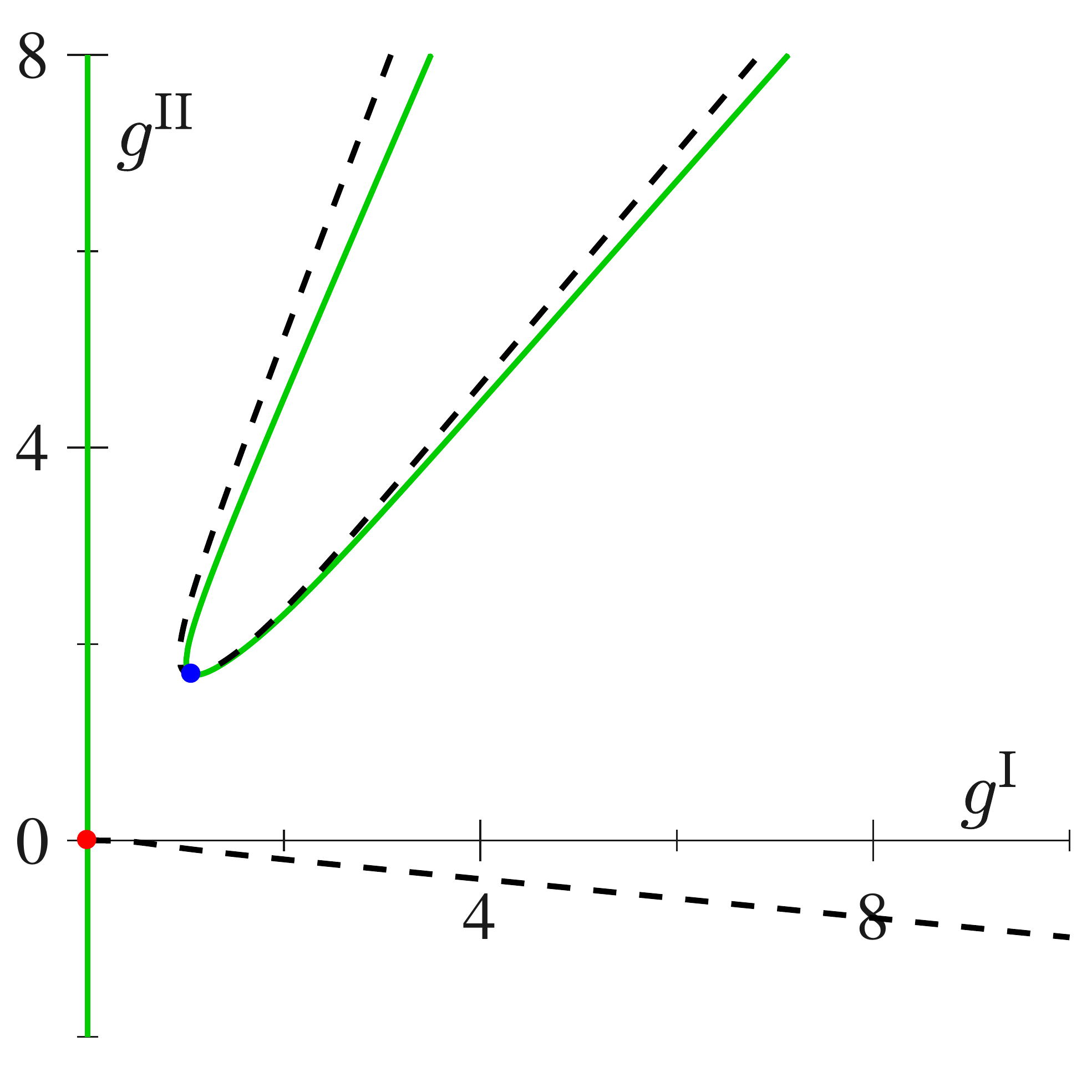}}
}
\caption{Zeros of $\beta^{\mathrm{I}}=0$ (green solid) and $\beta^{\mathrm{II}}=0$ (black dashed) for $d=3$. (a) At one-loop order, curves do not intersect except at the unstable Gaussian fixed point (red dot) at the origin. (b) At two-loop order, curves intersect at the stable strong-coupling fixed point (blue dot).}
\label{intersection}
\end{figure}

The critical RSB transitions in spin and structural glasses are universally signaled by the instability of the replicon fluctuations~\cite{AT78,CKPUZ14}. The critical field theory for this de Almeida-Thouless-Gardner universality class is governed by two cubic couplings, $g^{\T=\mathrm{I},\mathrm{II}}$. The beta functions, $\beta^{\T}\equiv \mu\frac{\partial g^{\T}}{\partial \mu}$, then dictate the RG flow for these couplings. At two-loop order with the minimal subtraction scheme~\cite{CY17}, we have
\begin{widetext}
\bea\label{beta1}
\beta^{\mathrm{I}}&=&\frac{(d-6)}{2}\gI-\frac{57}{16}\le(\gI\ri)^3+\frac{13}{2}\le(\gI\ri)^2\gII-\frac{11}{4}\gI\le(\gII\ri)^2\\
&&-\frac{42293}{2304}\le(\gI\ri)^5+\frac{35639}{576}\le(\gI\ri)^4\gII-\frac{22265}{288}\le(\gI\ri)^3\le(\gII\ri)^2+\frac{11987}{288}\le(\gI\ri)^2\le(\gII\ri)^3-\frac{1139}{144}\gI\le(\gII\ri)^4\, , \nonumber\\
\label{beta2}\beta^{\mathrm{II}}&=&\frac{(d-6)}{2}\gII-\frac{1}{8}\le(\gI\ri)^3-\frac{25}{16}\le(\gI\ri)^2\gII+\frac{7}{2}\gI\le(\gII\ri)^2-\frac{3}{2}\le(\gII\ri)^3\\
&&-\frac{571}{768}\le(\gI\ri)^5-\frac{11153}{2304}\le(\gI\ri)^4\gII+\frac{25615}{1152}\le(\gI\ri)^3\le(\gII\ri)^2-\frac{35879}{1152}\le(\gI\ri)^2\le(\gII\ri)^3+\frac{5099}{288}\gI\le(\gII\ri)^4-\frac{1931}{576}\le(\gII\ri)^5\, .\nonumber
\eea
\end{widetext}
The RG flow stops at points with $\beta^{\T}=0$, i.e., at fixed points.  Such points live at the intersections of curves on which $\beta^{\rm I}=0$ and those on which $\beta^{\rm II}=0$. At one-loop order for $d<\du$, these curves do not intersect except at the unstable Gaussian fixed point $g^{\T}=0$ [Fig.~\ref{intersection}(a)], but at two-loop order they do [Fig.~\ref{intersection}(b)]. This intersection results in a strong-coupling fixed point, visible only beyond one-loop order, just like the Caswell-Banks-Zaks fixed point in non-Abelian gauge theories~\cite{Caswell74,BZ82}.

\begin{figure*}[t]
\centerline{
\subfloat[$d=5.43$]{\includegraphics[width=0.25\textwidth]{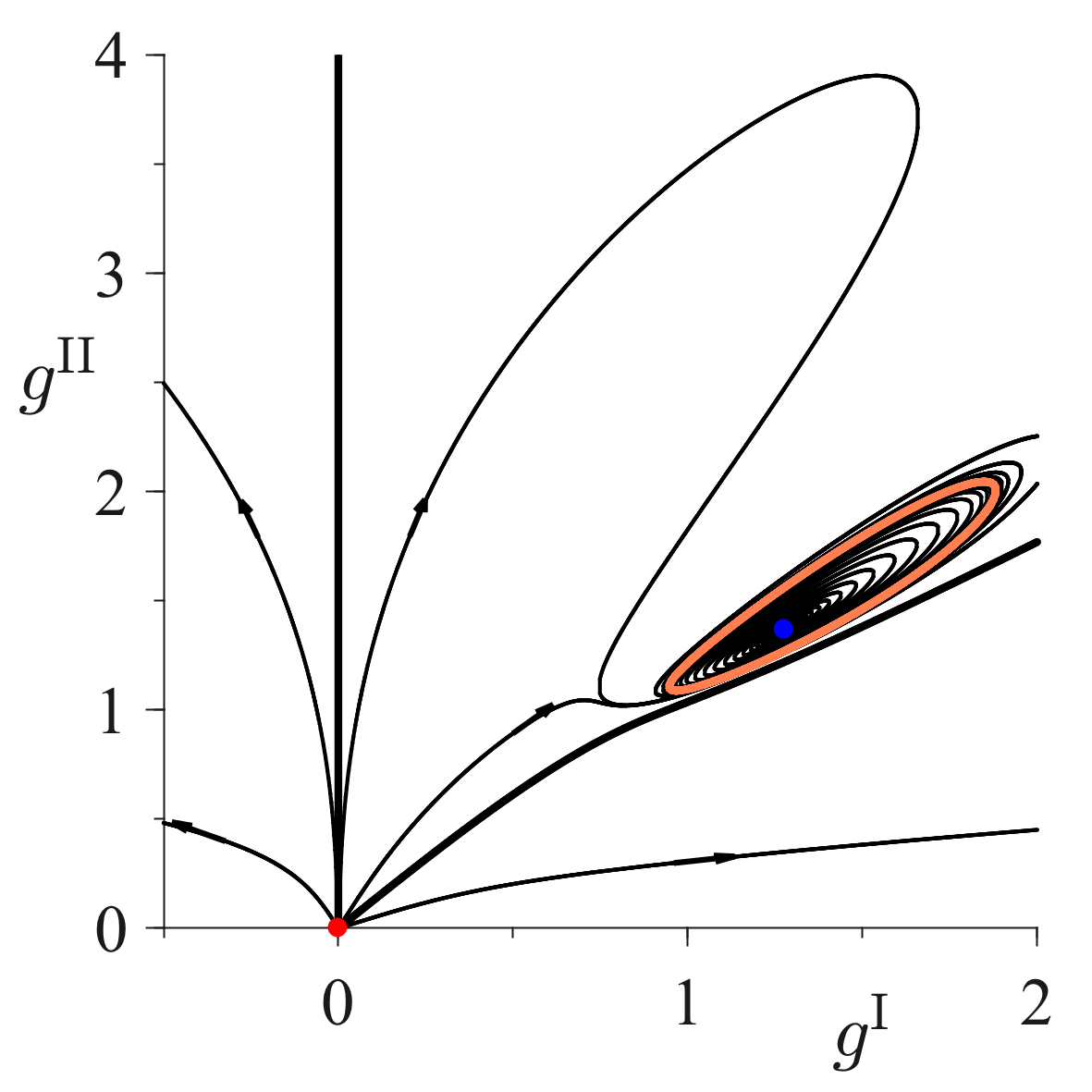}}\quad%
\hspace{-0.2in}
\subfloat[$d=5.50$]{\includegraphics[width=0.25\textwidth]{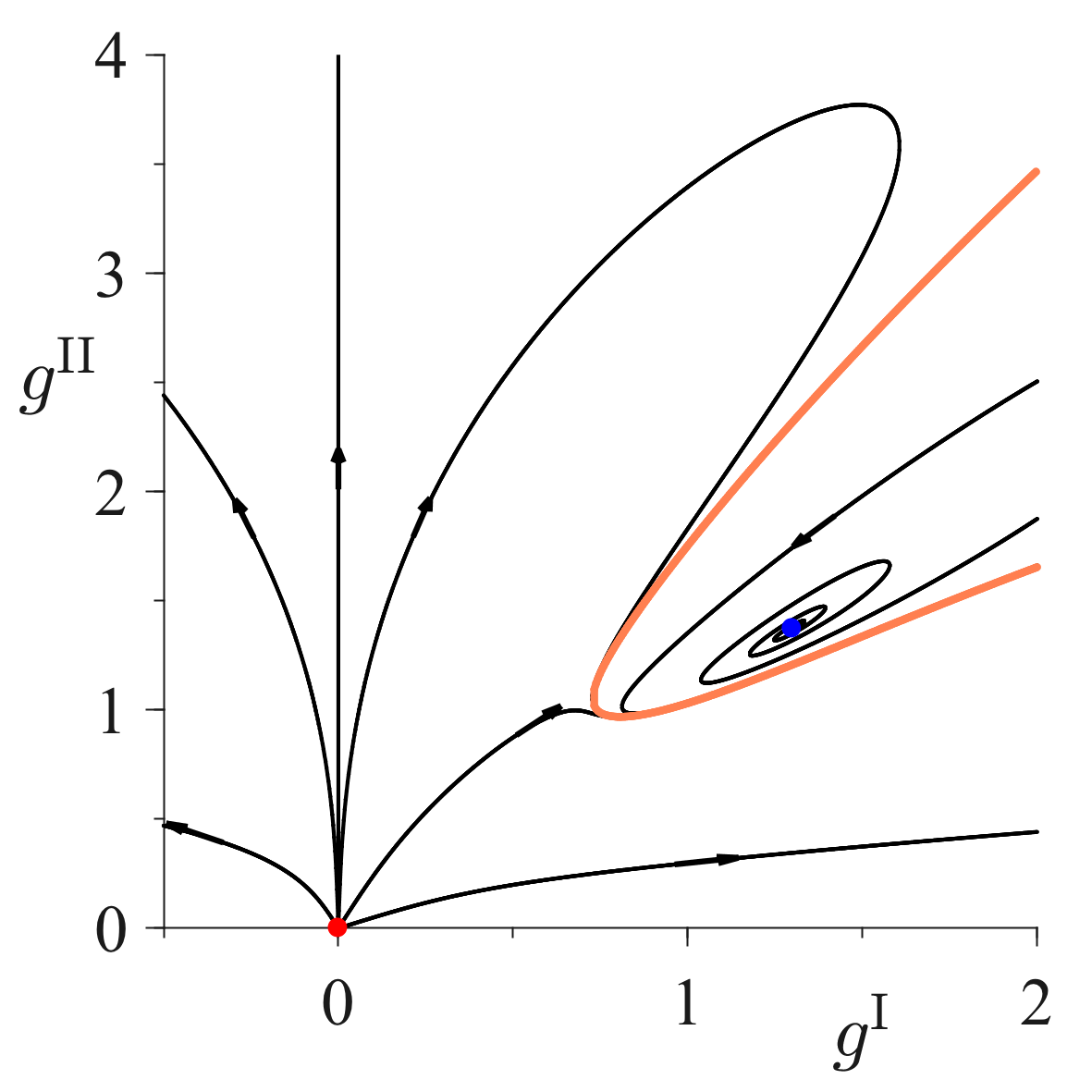}}\quad%
\hspace{-0.2in}
\subfloat[$d=6.005$]{\includegraphics[width=0.25\textwidth]{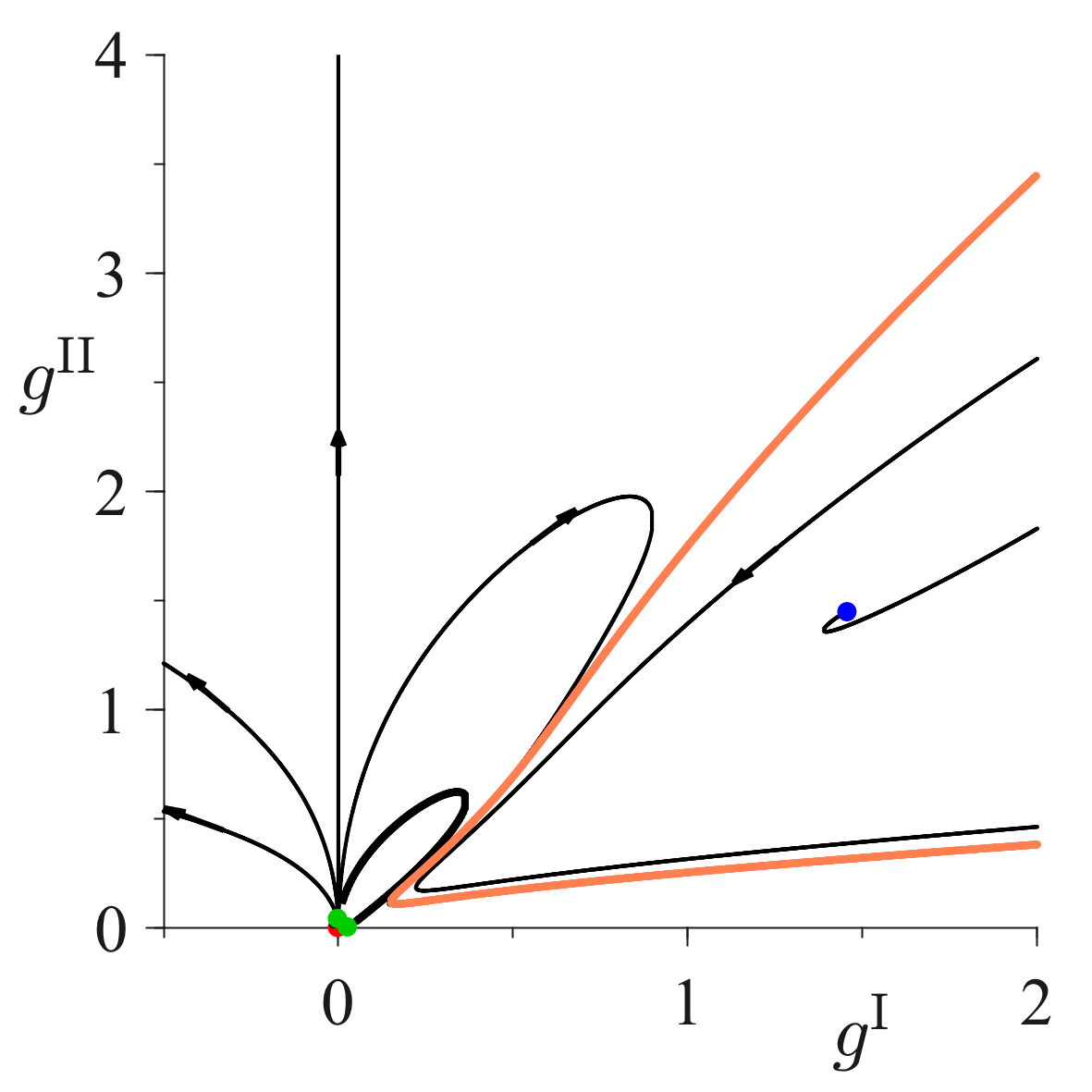}}\quad%
\hspace{-0.2in}
\subfloat[$d=6.05$]{\includegraphics[width=0.25\textwidth]{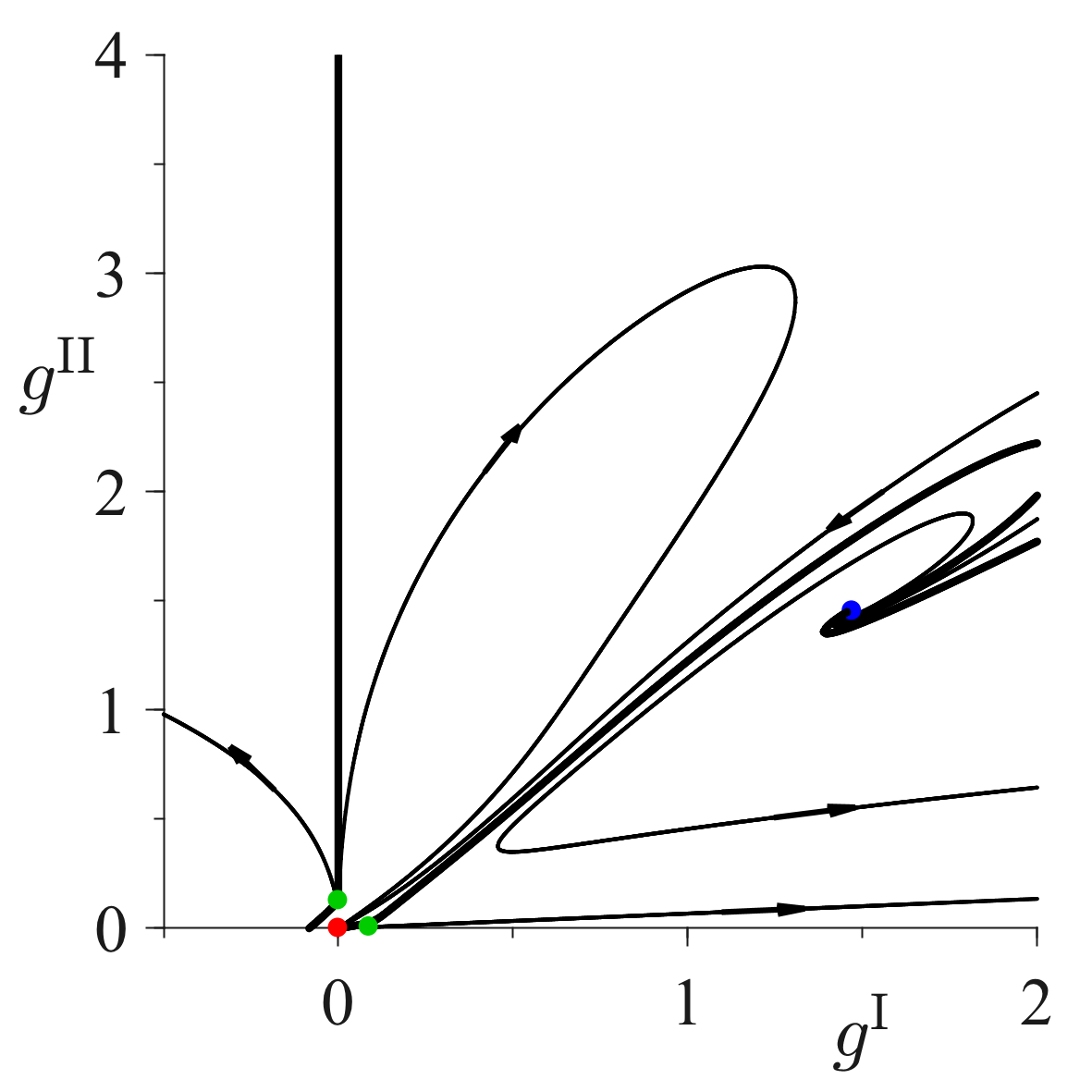}}
}
%\vspace{-1.86in}
%\centerline{\hspace{+8.5in}
%\includegraphics[width=0.284\textwidth]{color_strip.png}
%}
%\vspace{+0.25in}
\caption{Flows in the space of couplings for minimal two-loop RG equations.
(a) At $d=5.43\in(\dH,\dC)$, a limit cycle (thick orange line) is observed around the strong coupling fixed point (blue dot). Its basin of attraction is delineated by thick black lines.
(b) At $d=5.50\in[\dC,\du]$, an infinite-size remnant of the limit cycle is observed, but no stable fixed point. Within the remnant, the flow emanates from the strong-coupling fixed point and then circles around and approaches this infinite remnant.
(c) At $d=6.005\in(\du,\dcol)$, the Gaussian fixed point (red dot) is stable with a finite basin of attraction. On the boundary of the basin, two unstable fixed points (green dots) are observed along with their mirror images.
(d) At $d=6.05\geq\dcol$, the finite Gaussian basin and the infinite-size remnant of the limit cycle have merged. The Gaussian basin is now semi-infinite.
%Trajectories in this basin spiral out in a self-similar manner near the strong-coupling fixed point.
}
\label{minimalRGflow}
\end{figure*}

The flow geometry around the strong-coupling fixed point evolves with $d$. In order to study this dimensional dependence more carefully, we analyze the minimal two-loop RG flow equations~\eqref{beta1} and~\eqref{beta2} numerically~\cite{footnoteRK}.
%As a check we also obtained the same pictures through the pseudospectral method.
For $d\leq\dS\approx4.84$, the strong-coupling fixed point is stable. For $\dS<d<\dH\approx 5.41$, the fixed point is still stable but the stability exponents attain imaginary parts, causing the flow to spiral into the fixed point. At $d=\dH$, the real part of the stability exponents changes sign, making the fixed point unstable and resulting in the emergence of a stable limit cycle through a Hopf bifurcation [Fig.~\ref{minimalRGflow}(a)]. Such a limit cycle gives rise to a log-periodic, discrete scale invariance in physical observables. Such scale invariance is familiar from the period-doubling route to chaos, and is speculated to be important in stock market crashes~\cite{SornetteJB96}, earthquakes, and many other systems~\cite{Sornette98}. Our results indicate that spin and structural glasses might therefore share a 
%deep
connection with these phenomena.

The size of this stable limit cycle cascades toward infinity as dimension nears $\dC\approx5.47$,
% $dC\approx 5.4700385$
leaving its infinite remnant [Fig.~\ref{minimalRGflow}(b)].
%with possible Landau pole at infinity.
For $d\in[\dC,\du]$, there is neither a stable fixed point nor a finite stable limit cycle in sight of the minimal two-loop analysis. Analytically continuing the flow equations above $\du$,
%--the legitimacy of which is highly questionable but the blind application of which is highly fashionable--
the Gaussian fixed point becomes stable with a finite basin of attraction of size $\propto \sqrt{d-\du}$ [Fig.~\ref{minimalRGflow}(c)], and at $d=\dcol\approx 6.01$ this basin collides with the infinite remnant of the limit cycle discussed above, resulting in a semi-infinite basin of attraction for the Gaussian fixed point [Fig.~\ref{minimalRGflow}(d)].

In Ref.~\cite{MR18}, through the analysis of the one-loop RG flow, Moore and Read predicted the existence of a multicritical point --- and of a nonperturbative phase transition of an indeterminate kind --- on the de Almeida-Thouless line. Their argument, which is based on the shrinkage of the basin of attraction as $d\rightarrow\du^{+}$ and the absence of the critical fixed point for $d<\du$ in the weak-coupling regime, still applies to the minimal two-loop RG flow in the window $d\in[\dC,\dcol)\approx[5.41,6.01)$.
In the next section, we suggest an alternative scenario that emerges upon transforming coordinates of two-loop RG equations.

%\section{Dependence on coordinate transformations: The normal form}
\section{Dependence on coordinate transformations}
\label{normal}
In Ref.~\cite{CY17}, a three-loop calculation with Borel resummation was performed to further corroborate the existence of the critical fixed point identified at two-loop order. However, the resummation scheme employed was admittedly ad hoc, partly due to scarcity of systematic studies on resummation schemes for field theories with two couplings (see, however, Ref.~\cite{Dunne05}). Nonperturbative RG equations also exist for this problem (see Appendix~\ref{NRG}). Although a partial analysis suggests that their predictions are consistent with those of the two-loop analysis, the results also suffer from the uncontrolled scheme dependence. Just as for the Ising universality class, the existence and nature of the fixed point in turn depend on the details of the scheme used. 
%Another thing we tried is to mock with coefficients, and we could in particular make the limit cycle survive all the way up to $d=\du$. E.g., multiplying the coefficient of $\gI^0 \gII^3$ in $\beta^{\mathrm{II}}$ by $s=0.7$, the fixed point becomes even stable (i.e., the size of limit cycle is zero) at $d=\du$. (Even though this way of mocking is affecting one-loop coefficient, the one-loop topology remains the same.) To do this less disturbingly, we could modify a coefficients of two-loop part, not one-loop part. E.g., we have been able to do this by multiplying the coefficient of  $\gI^0 \gII^5$ in $\beta^{\mathrm{II}}$, though in this case I had to make $s\approx0.99$ and making it too small would make the fixed point die.
The coordinate-transformation scheme we present below is no exception to this lack of systematics. It nonetheless yields a simple scenario, in which the critical RSB transition survives for all spatial dimensions $d$. The proposal should thus be of interest for the community to keep in mind.

Generically, external parameters controlled in experiments and simulations map nontrivially to coupling coordinates of effective field theories in the RG analysis. It is therefore natural to analyze the dependence of the fixed point structure against changes in coupling coordinates. In particular universal properties near a fixed point should be invariant under coordinate changes, and this invariance can be used to cast the RG equations into a normal form around that fixed point~\cite{RCHKLRS17}. Below, we perform a coordinate change around the Gaussian fixed point and explore its effect on the strong-coupling fixed point.

Scrutinizing the structure of Feynman diagrams lets us organize the perturbative RG flow equation into the form~\cite{CY17}
\bea\label{2loopFeynman}
\mu\frac{\partial g^{\T}}{\partial \mu}&=&\le[-\frac{\epsilon}{2}+\frac{1}{4}I_2(g)-\frac{11}{144}I_2^2(g)+\frac{1}{6}I_4(g)\ri]g^{\T}\, \\
&&-I_{3}^{\T}(g)+\frac{7}{24}I_2(g)I_3^{\T}(g)-\frac{1}{2}I_{5,A}^{\T}(g)-\frac{3}{4}I_{5,B}^{\T}(g) \nonumber\, 
\eea
where $\epsilon\equiv\du-d$ and $I_k(g)$'s are $k$-th degree homogeneous polynomials of two variables $g^{\T=\mathrm{I},\mathrm{II}}$ (see Appendix~\ref{CT}). We keep this algebraic structure suggested by the Feynman diagrams intact and thus restrict ourselves to the class of coordinate transformations involving only these polynomials.
% Some of the ``asymptoticness" in the original series may be absorbed into the ``asymptoticness" of the coordinate transformation. Even if there is a linear growth of the number of coefficients for two-coupling theories, can we always make beta-functions analytic, absorbing all the asymptoticness into the coordinate transformation? Feynman-diagrammatic-way of understanding a ``good" normal-coordinate choice?
Specifically, we recast the RG-flow equations in a new normal-form coordinate $\tilde{g}^{\T}$ defined through
\begin{widetext}
\be
g^{\T}=\tilde{g}^{\T}+\lambda_1\tilde{g}^{\T} I_2(\tilde{g})+\lambda_2 I_{3}^{\T}(\tilde{g})+\Lambda_1\tilde{g}^{\T} I_2^2(\tilde{g})+\Lambda_2\tilde{g}^{\T} I_4(\tilde{g})+\Lambda_3 I_2(\tilde{g})I_3^{\T}(\tilde{g})+\Lambda_4 I_{5,A}^{\T}(\tilde{g})+\Lambda_5 I_{5,B}^{\T}(\tilde{g})+O(\tilde{g}^7)
\ee
%\end{widetext}
and truncate higher-order terms. After some algebra we obtain
\bea\label{betatilde1}
%\eta&=&\frac{1}{6}I_2(\tilde{g})+\le(-\frac{11}{216}+\frac{1}{3}\lambda_1\ri)I_2^2(\tilde{g})+\le(\frac{1}{9}+\frac{1}{3}\lambda_2\ri)I_4(\tilde{g})+O(\tilde{g}^6)\,, \\
%\nu^{-1}&=&2-\eta+I_2(\tilde{g})+\le(-\frac{1}{24}+2\lambda_1\ri)I_2^2(\tilde{g})+\le(1+2\lambda_2\ri)I_4(\tilde{g})+O(\tilde{g}^6)\, \\
&&\mu\frac{\partial \tilde{g}^{\T}}{\partial\mu}\\
&=&\le\{-\frac{\epsilon}{2}+\le(\frac{1}{4}+\epsilon\lambda_1\ri)I_2(\tilde{g})+\le[-\frac{11}{144}+\epsilon\le(-3\lambda_1^2+2\Lambda_1\ri)\ri]I_2^2(\tilde{g})+\le[\frac{1}{6}+2\lambda_1+\frac{1}{2}\lambda_2+\epsilon\le(-2\lambda_1\lambda_2+2\Lambda_2\ri)\ri]I_4(\tilde{g})\ri\}\tilde{g}^{\T}\, \nonumber\\
&&+(-1+\epsilon\lambda_2)I_{3}^{\T}(\tilde{g})+\le[\frac{7}{24}-2\lambda_1-\frac{1}{2}\lambda_2+\epsilon\le(-4\lambda_1\lambda_2+2\Lambda_3\ri)\ri]I_2(\tilde{g})I_3^{\T}(\tilde{g})\, \nonumber\\
&&+\le(-\frac{1}{2}+2\epsilon\Lambda_4\ri)I_{5,A}^{\T}(\tilde{g})+\le[-\frac{3}{4}+\epsilon\le(-3\lambda_2^2+2\Lambda_5\ri)\ri]I_{5,B}^{\T}(\tilde{g})+O(\tilde{g}^7) \, .\nonumber
\eea
Here, we choose $\Lambda_{1,2,3,4,5}$ appropriately to cancel the $\epsilon$-dependent quintic terms, which yields
\bea
\mu\frac{\partial \tilde{g}^{\T}}{\partial\mu}&=&\le[-\frac{\epsilon}{2}+\le(\frac{1}{4}+\epsilon\lambda_1\ri)I_2(\tilde{g})-\frac{11}{144}I_2^2(\tilde{g})+\le(\frac{1}{6}+2\lambda_1+\frac{1}{2}\lambda_2\ri)I_4(\tilde{g})\ri]\tilde{g}^{\T}\, \\
&&+(-1+\epsilon\lambda_2)I_{3}^{\T}(\tilde{g})+\le(\frac{7}{24}-2\lambda_1-\frac{1}{2}\lambda_2\ri)I_2(\tilde{g})I_3^{\T}(\tilde{g})-\frac{1}{2}I_{5,A}^{\T}(\tilde{g})-\frac{3}{4}I_{5,B}^{\T}(\tilde{g})+O(\tilde{g}^7) \, .\nonumber
\eea

For $d=\du$ the flow equation depends only on one parameter, the linear combination $\lambda\equiv-2\lambda_1-\frac{1}{2}\lambda_2$:
%any rationale for this ``gauge redundancy"?
%\begin{widetext}
\be\label{betatilde2}
\mu\frac{\partial \tilde{g}^{\T}}{\partial\mu}=\le[\frac{1}{4}I_2(\tilde{g})-\frac{11}{144}I_2^2(\tilde{g})+\le(\frac{1}{6}-\lambda\ri)I_4(\tilde{g})\ri]\tilde{g}^{\T}-I_{3}^{\T}(\tilde{g})+\le(\frac{7}{24}+\lambda\ri)I_2(\tilde{g})I_3^{\T}(\tilde{g})-\frac{1}{2}I_{5,A}^{\T}(\tilde{g})-\frac{3}{4}I_{5,B}^{\T}(\tilde{g}) \, .
\ee
\end{widetext}
The existence of the strong-coupling fixed point is robust against $\lambda$-deformation within the window $\lambda\in[-0.91,1.19]$. In addition, the fixed point becomes stable for $\lambda>\lamAH\approx1.00$ through an anti-Hopf bifurcation. In other words, for $\lambda<\lamAH$ the fixed point is unstable without a limit cycle around it, while for $\lambda>\lamAH$ it is stable with an unstable limit cycle around it.

\begin{figure}[t]
\centerline{
\subfloat[$d=3$]{\includegraphics[width=0.48\textwidth]{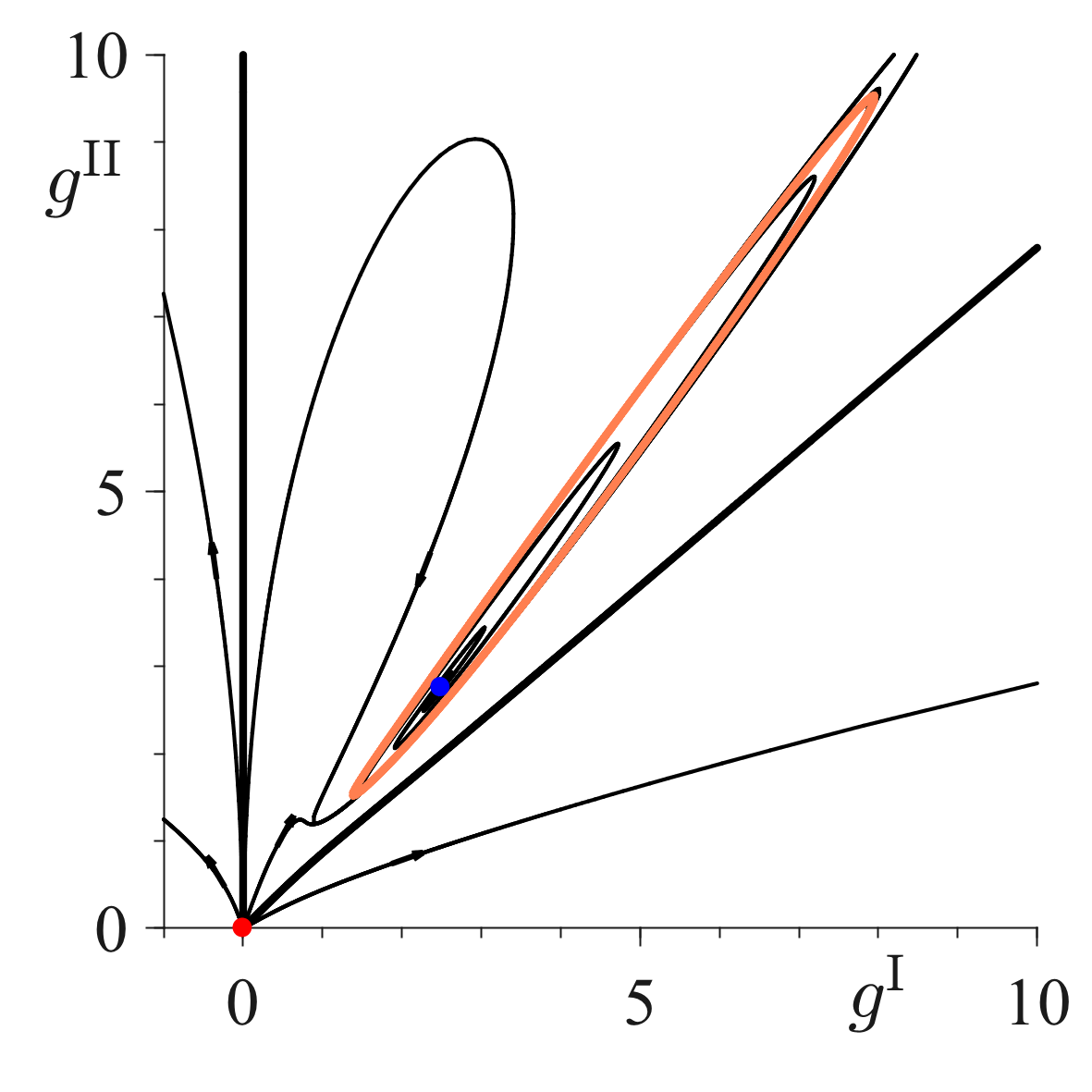}}\quad
\subfloat[$d=6.005$]{\includegraphics[width=0.48\textwidth]{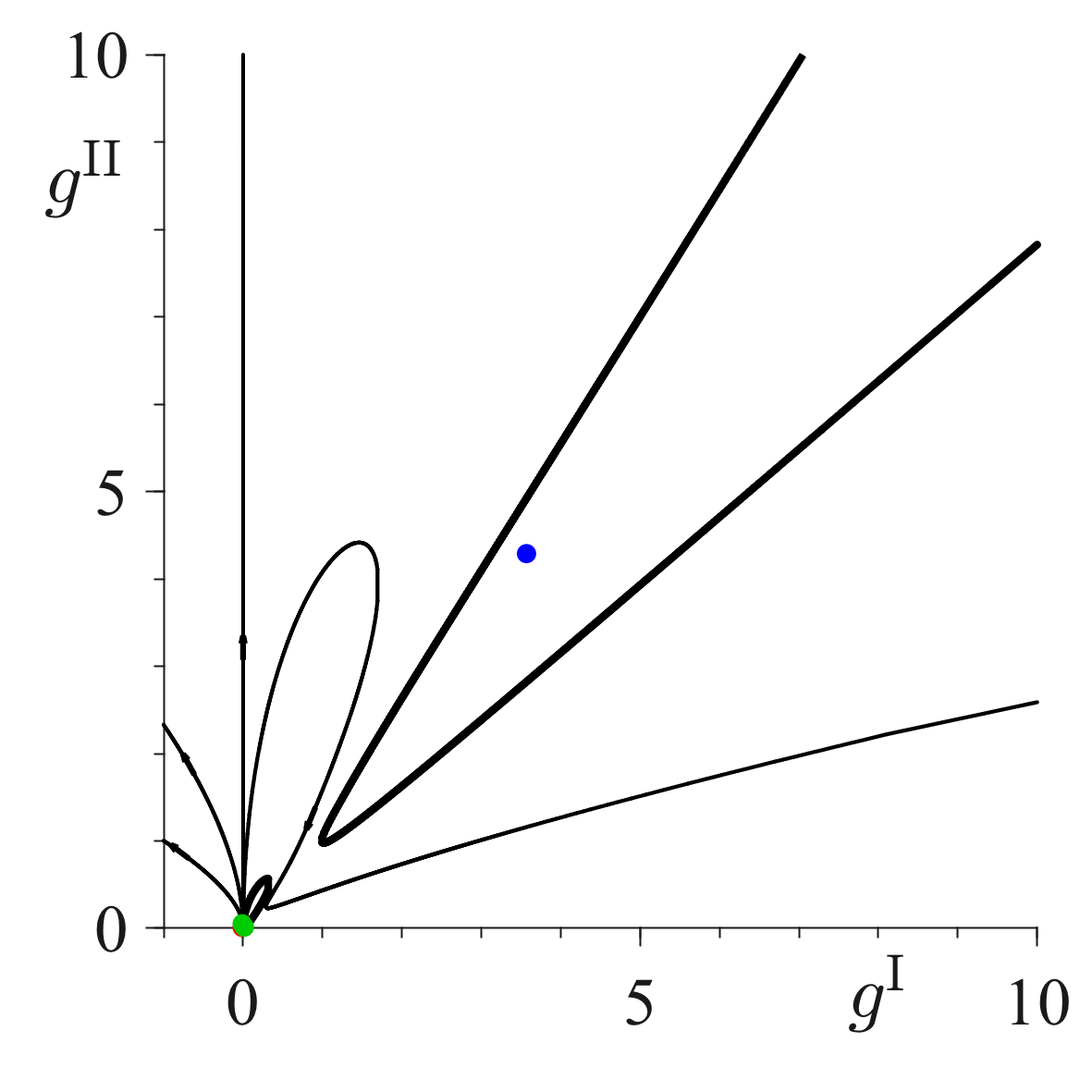}}
}
\caption{Flows in the space of couplings for coordinate-transformed two-loop RG scheme, with $\lambda_1=-0.55$ and $\lambda_2=0$. 
(a) At $d=3$, a limit cycle (thick orange line) is observed around the strong coupling fixed point. Its basin of attraction is delineated by thick black lines.
(b) At $d=6.005$, two basins of attraction are observed: one for the Gaussian fixed point and the other for the strong-coupling fixed point.}
\label{ct2RG}
\end{figure}

For $d\ne\du$, the space of coordinate changes is two dimensional. While this is much simpler than the full space of coordinate changes (recall that we chose to respect the algebraic structure mentioned above and in addition chose to cancel $\epsilon$-dependences of the highest order terms in the RG equations), there are still a myriad of possibilities depending on values of $\lambda_{1,2}$ ($\lambda_{1,2}$ can even be dependent on $d$).
% as long as analytic). 
We will not thoroughly investigate them all because, without a guiding principle to dictate the desired properties of the coordinate transformation, such effort would be mostly moot. Instead, below we illustrate one physical scenario given by the choice $\lambda_1=-0.55$ and $\lambda_2=0$. For this choice and $d$ just above $\du$, two basins of attractions can be found: one for the Gaussian fixed point and the other for the strong-coupling fixed point [Fig.~\ref{ct2RG}(b)]. There, depending on the microscopic details of the model, the de Almeida-Thouless-Gardner critical line may then do one of the following: (i) lie completely within the Gaussian basin, in which case one observes the mean-field criticality; (ii) lie completely within the strong-coupling basin, in which case one observes non-mean-field criticality; (iii) cross borders of basins, in which case the line fragments into several parts; or (iv) not lie within any basin, in which case one might not observe criticality.

As $d\rightarrow\du^{+}$, the Gaussian basin shrinks to zero while the strong-coupling basin remains nonzero. Upon further decreasing $d$, the strong-coupling fixed point goes through a Hopf bifurcation at $d\approx4.87$, below which it has a stable limit cycle around it [Fig.~\ref{ct2RG}(a)]. Note that this process is the opposite of what happens within the minimal two-loop RG scheme, in which the Hopf bifurcation results in the limit cycle upon increasing $d$. This specific scheme results in discrete scale invariance instead being observed in low dimensions. Interestingly, a nontrivial critical fixed point in high dimension was also found in Ref.~\cite{AB15},
% one for dAT, another for $h\rightarrow0$ for sufficiently low temperatures
within a Migdal-Kadanoff RG scheme, but no critical limit cycle was then found upon lowering dimensions.

In summary, with this choice of transformation, the basin of attraction for the strong-coupling criticality stays nonzero and, if a given model lies within it for all $d$, a dimensionally-robust nontrivial criticality is expected.

\section{Conclusion}
\label{conclusion}
We have analyzed the higher-loop RG flow equations to explore scenarios that are invisible at one-loop order. The analysis of the minimal two-loop RG flow equations reveals a strong-coupling critical fixed point, as first reported in Ref.~\cite{CY17}, and a more careful analysis of their dimensional dependence discloses a critical limit cycle.
We have additionally explored the challenge of extrapolating perturbative RG calculations far from the Gaussian fixed point, especially through their dependence on the choice of perturbative coordinate changes in coupling space. For the critical RSB field theory analyzed herein, such coordinate transformations on the two-loop equation depict several plausible physical predictions, one of which suggests that the basin of attraction of the critical RSB transition stays nonzero in all spatial dimensions $d$, with a limit cycle in lower dimensions. These scheme dependencies highlight the need for further development in resummation, coordinate-transformed, and nonperturbative RG schemes. The critical RSB field theory should serve as a crucial testing ground for these advances.

In addition to persistent theoretical investigations, experiments and simulations on a diverse set of systems will be indispensable to determine the role of RSB transitions in finite dimensions. To emphasize this point, let us imagine a given model that lies close to the Gaussian fixed point. In that case, even if a strong-coupling fixed point exists, the one-loop scenario would still apply, with the the de Almeida-Thouless line fragmenting upon $d\rightarrow\du$, as proposed in Ref.~\cite{MR18} and criticality being absent below $\du$. It is also possible that a given model might stay outside the basins of the critical fixed points, in which case it would not exhibit any sign of criticality, just as in the droplet scenario. These considerations show that an absence of RSB criticality in a few model systems may be due to their unfortunate locations in coupling spaces and cannot be invoked to exclude the presence of criticality in other systems. By contrast, a single observation of RSB criticality for $d<\du$ would indicate the existence of a nontrivial critical fixed point. In particular, if discrete scale invariance were observed in any dimension, it would substantially support the strong-coupling criticality scenario proposed in Ref.~\cite{CY17} and herein.

\begin{acknowledgments}
We thank Giulio Biroli and Michael A.~Moore for discussions.
P.~C., Y.~H., and S.~Y.~acknowledge support from the Simons Foundation grant (\#454937, Patrick Charbonneau) and A.~R.~and J.~P.~S.~acknowledge support from the National Science Foundation Grant No. NSF DMR-1719490.
%Data relevant to this work have been archived and can be accessed at http://dx.doi.org/??.????/????????.
\end{acknowledgments}

\clearpage

\appendix
\renewcommand{\thefigure}{S\arabic{figure}}
\setcounter{figure}{0}
\begin{widetext}

\section{Combinatorial factors}
\label{CT}
As discussed in Ref.~\cite{CY17}, the critical replicon field, $\phi_{ab}\le(\mathbf{x}\ri)$, is symmetric, \textit{i.e.}, $\phi_{ab}=\phi_{ba}$ for replica indices $a,b$ running from $1$ to $n$, has no diagonal degree of freedom, \textit{i.e.}, $\phi_{aa}=0$, and further satisfies the replicon conditions $\sum_{b=1}^{n} \phi_{ab}=0$.
By defining an orthonormal basis $\le\{e^{i}_{ab}\ri\}_{i=1,\ldots,n(n-3)/2}$ through
\be
\sum_{a,b=1}^{n} e^{i}_{ab} e^{j}_{ab}=\delta^{ij}\, ,
\ee
\be
e^{i}_{aa}=0\, ,
\ee
and
\be
\sum_{b=1}^{n} e^{i}_{ab}=0\, 
\ee
for all $a=1,\ldots,n$, we can expand the replicon field as
\be
\phi_{ab}\le(\mathbf{x}\ri)=\sum_{i=1}^{\frac{n(n-3)}{2}}\phi_i\le(\mathbf{x}\ri) e^{i}_{ab}\, .
\ee
The critical Lagrangian can then be expressed as
\bea
\mathcal{L}&=&\frac{1}{2}\sum_{a,b=1}^{n}\le(\nabla\phi_{ab}\ri)^2-\frac{1}{3!}\le(g^{\rm I}\sum_{a,b=1}^n\phi_{ab}^3+g^{\rm II}\sum_{a,b,c=1}^n\phi_{ab}\phi_{bc}\phi_{ca}\ri)\nonumber\\
&=&\frac{1}{2}\sum_{i=1}^{\frac{n(n-3)}{2}}\le(\nabla\phi_i\ri)^2-\frac{1}{3!}\sum_{i,j,k=1}^{\frac{n(n-3)}{2}}\le(g^{\rm I}T_{\rm I}^{ijk}+g^{\rm II}T_{\rm II}^{ijk}\ri)\phi_{i}\phi_{j}\phi_{k}
\eea
with
\be
T_{\rm I}^{ijk}\equiv \sum_{a,b=1}^{n}e^{i}_{ab}e^{j}_{ab}e^{k}_{ab}
\ee
and
\be
T_{\rm II}^{ijk}\equiv \sum_{a,b,c=1}^{n}e^{i}_{ab}e^{j}_{bc}e^{k}_{ca}\, .
\ee

The homogeneous polynomials that appear in Eq.~\eqref{2loopFeynman} are defined as
\bea
I_2(g)&\equiv&\sum_{\T_1,\T_2\in\le\{\mathrm{I},\mathrm{II}\ri\}}S_{\T_1,\T_2}g^{\T_1}g^{\T_2}\, ,\label{simple1}\\
I^{\T}_{3}(g)&\equiv&\sum_{\T_1,\T_2,\T_3\in\le\{\mathrm{I},\mathrm{II}\ri\}}a^{\T}_{\T_1,\T_2,\T_3}g^{\T_1}g^{\T_2}g^{\T_3}\, ,\\
I_{4}(g)&\equiv&\sum_{\T_1,\T_2,\T_3,\T_4,\T_5\in\le\{\mathrm{I},\mathrm{II}\ri\}}S_{\T_1,\T_5}a^{\T_5}_{\T_2,\T_3,\T_4}g^{\T_1}g^{\T_2}g^{\T_3}g^{\T_4}\, ,\\
I^{\T}_{5,A}(g)&\equiv&\sum_{\T_1,\T_2,\T_3,\T_4,\T_5\in\le\{\mathrm{I},\mathrm{II}\ri\}}a^{\T}_{\T_1,\T_2,\T_3;\T_4,\T_5}g^{\T_1}g^{\T_2}g^{\T_3}g^{\T_4}g^{\T_5}\, ,\\
I^{\T}_{5,B}(g)&\equiv&\sum_{\T_1,\T_2,\T_3,\T_4,\T_5,\T_6\in\le\{\mathrm{I},\mathrm{II}\ri\}}a^{\T}_{\T_1,\T_2,\T_6}a^{\T_6}_{\T_3,\T_4,\T_5}g^{\T_1}g^{\T_2}g^{\T_3}g^{\T_4}g^{\T_5}\, .\label{simple5}
\eea
The one-loop self-energy combinatorial factors, defined through
\be\label{defineS2}
\sum_{i_3,i_4=1}^{\frac{n(n-3)}{2}}T_{\T_1}^{i_1 i_3 i_4}T_{\T_2}^{i_2 i_4 i_3}=S_{\T_1, \T_2} \delta^{i_1i_2}\, ,
\ee
satisfies $S_{\T_1, \T_2}=S_{\T_2, \T_1}$, the one-loop cubic factors $a^{\T}_{\T_1,\T_2,\T_3}$, defined through
\be\label{definea3}
\sum_{i_4,i_5,i_6=1}^{\frac{n(n-3)}{2}}T_{\T_1}^{i_1 i_5 i_6}T_{\T_2}^{i_2 i_6 i_4}T_{\T_3}^{i_3 i_4 i_5} =\sum_{\T\in\le\{\mathrm{I},\mathrm{II}\ri\}}a^{\T}_{\T_1,\T_2,\T_3}T_{\T}^{i_1 i_2 i_3}\, ,
\ee
are symmetric under permutations of indices $(\T_1,\T_2,\T_3)$, and the two-loop cubic factors $a^{\T}_{\T_1,\T_2,\T_3;\T_4,\T_5}$, defined through
\be
\label{definea5}
\sum_{i_4,i_5,i_6,i_7,i_8,i_9=1}^{\frac{n(n-3)}{2}}T_{\T_1}^{i_1 i_5 i_6}T_{\T_2}^{i_2 i_4 i_8}T_{\T_3}^{i_3 i_7 i_9}T_{\T_4}^{i_4 i_6 i_9}T_{\T_5}^{i_5 i_7 i_8}
\equiv \sum_{\T\in\le\{\mathrm{I},\mathrm{II}\ri\}}a^{\T}_{\T_1,\T_2,\T_3,\T_4,\T_5} T_{\T}^{i_1 i_2 i_3}\, ,
\ee
are symmetric under permutations of the first three indices $(\T_1,\T_2,\T_3)$ and of the last two indices $(\T_4,\T_5)$. Explicitly, these combinatorial factors are given by
\be\label{Ss}
\left[ {\begin{array}{c}
   S_{{\rm I},{\rm I}}  \\
   S_{{\rm I},{\rm II}}  \\
   S_{{\rm II},{\rm II}}   \\
  \end{array} } \right]=\left[ {\begin{array}{c}
   \frac{n^3-9n^2+26n-22}{2(n-1)(n-2)^2}  \\
   \frac{3n^2-15n+16}{2(n-1)(n-2)^2}  \\
   \frac{n^4-8n^3+19n^2-4n-16}{4(n-1)(n-2)^2}   \\
  \end{array} } \right]\, ,
\ee
\be\label{a3s}
\left[ {\begin{array}{cc}
   a^{\mathrm{I}}_{{\rm I},{\rm I},{\rm I}} & a^{\mathrm{II}}_{{\rm I},{\rm I},{\rm I}} \\
   a^{\mathrm{I}}_{{\rm I},{\rm I},{\rm II}} & a^{\mathrm{II}}_{{\rm I},{\rm I},{\rm II}} \\
  a^{\mathrm{I}}_{{\rm I},{\rm II},{\rm II}}  & a^{\mathrm{II}}_{{\rm I},{\rm II},{\rm II}} \\
  a^{\mathrm{I}}_{{\rm II},{\rm II},{\rm II}} & a^{\mathrm{II}}_{{\rm II},{\rm II},{\rm II}} \\
  \end{array} } \right]=\left[ {\begin{array}{cc}
    \frac{n^3-11n^2+38n-34}{2(n-1)(n-2)^2} & \frac{-1}{(n-2)^3} \\
    \frac{3n^2-19n+20}{2(n-1)(n-2)^2}  & \frac{-n^3+8n^2-17n+12}{2(n-1)(n-2)^3} \\
   \frac{-n^3+5n^2+8n-16}{4(n-1)(n-2)^2}  & \frac{3n^3-27n^2+64n-48}{4(n-1)(n-2)^3} \\
   \frac{-3n}{2(n-2)^2}  & \frac{n^5-10n^4+33n^3-8n^2-104n+112}{8(n-1)(n-2)^3} \\
  \end{array} } \right]\, ,
\ee
\be
\label{a5sI}
\left[ {\begin{array}{c}
  a^{\mathrm{I}}_{{\rm I},{\rm I},{\rm I};{\rm I},{\rm I}}  \\
  a^{\mathrm{I}}_{{\rm II},{\rm I},{\rm I};{\rm I},{\rm I}} \\
  a^{\mathrm{I}}_{{\rm I},{\rm I},{\rm I};{\rm II},{\rm I}}  \\
  a^{\mathrm{I}}_{{\rm I},{\rm I},{\rm I};{\rm II},{\rm II}}  \\
  a^{\mathrm{I}}_{{\rm II},{\rm II},{\rm I};{\rm I},{\rm I}} \\
  a^{\mathrm{I}}_{{\rm II},{\rm I},{\rm I};{\rm II},{\rm I}}  \\
  a^{\mathrm{I}}_{{\rm II},{\rm II},{\rm II};{\rm I},{\rm I}}  \\
  a^{\mathrm{I}}_{{\rm I},{\rm I},{\rm II};{\rm II},{\rm II}}  \\
  a^{\mathrm{I}}_{{\rm I},{\rm II},{\rm II};{\rm I},{\rm II}} \\
  a^{\mathrm{I}}_{{\rm I},{\rm II},{\rm II};{\rm II},{\rm II}}  \\
  a^{\mathrm{I}}_{{\rm II},{\rm II},{\rm II};{\rm I},{\rm II}}  \\  
  a^{\mathrm{I}}_{{\rm II},{\rm II},{\rm II};{\rm II},{\rm II}} \\
  \end{array} } \right]
  \equiv
  \left[ {\begin{array}{c}
  \frac{n^8-26n^7+291n^6-1816n^5+6840n^4-15756n^3+21586n^2-16088n+5008}{4(n-1)^2 (n-2)^6}  \\
  \frac{3n^7-66n^6+607n^5-2960n^4+8132n^3-12592n^2+10236n-3392}{4(n-1)^2 (n-2)^6}  \\
  \frac{3n^7-66n^6+604n^5-2930n^4+8017n^3-12380n^2+10048n-3328}{4(n-1)^2 (n-2)^6}  \\
  \frac{21n^6-366n^5+2493n^4-8316n^3+14536n^2-12800n+4480}{8(n-1)^2 (n-2)^6}  \\
  \frac{3n^7-27n^6-59n^5+1471n^4-6396n^3+12496n^2-11664n+4224}{8(n-1)^2 (n-2)^6}  \\
  \frac{n^7-7n^6-63n^5+819n^4-3292n^3+6262n^2-5776n+2080}{4(n-1)^2 (n-2)^6}  \\
  \frac{n^9-19n^8+145n^7-541n^6+1018n^5-1488n^4+4292n^3-10192n^2+11328n-4608}{16(n-1)^2 (n-2)^6} \\
  \frac{-n^7+20n^6-110n^5+84n^4+871n^3-2704n^2+3040n-1216}{4(n-1)^2 (n-2)^6}  \\
  \frac{-7n^7+134n^6-819n^5+1708n^4+680n^3-7552n^2+10144n-4352}{16(n-1)^2 (n-2)^6}  \\
  \frac{n^9-15n^8+95n^7-469n^6+2196n^5-6368n^4+8592n^3-2176n^2-5376n+3584}{32(n-1)^2 (n-2)^6}  \\
  \frac{3n^8-42n^7+169n^6+68n^5-1750n^4+3488n^3-1456n^2-1984n+1536}{16(n-1)^2 (n-2)^6}  \\
  \frac{n(-3n^6+54n^5-315n^4+560n^3+376n^2-1968n+1440)}{16(n-1)(n-2)^6}  \\
    \end{array} } \right]\, ,
\ee
and
\be
\label{a5sII}
\left[ {\begin{array}{c}
  a^{\mathrm{II}}_{{\rm I},{\rm I},{\rm I};{\rm I},{\rm I}}  \\
  a^{\mathrm{II}}_{{\rm II},{\rm I},{\rm I};{\rm I},{\rm I}} \\
  a^{\mathrm{II}}_{{\rm I},{\rm I},{\rm I};{\rm II},{\rm I}}  \\
  a^{\mathrm{II}}_{{\rm I},{\rm I},{\rm I};{\rm II},{\rm II}}  \\
  a^{\mathrm{II}}_{{\rm II},{\rm II},{\rm I};{\rm I},{\rm I}} \\
  a^{\mathrm{II}}_{{\rm II},{\rm I},{\rm I};{\rm II},{\rm I}}  \\
  a^{\mathrm{II}}_{{\rm II},{\rm II},{\rm II};{\rm I},{\rm I}}  \\
  a^{\mathrm{II}}_{{\rm I},{\rm I},{\rm II};{\rm II},{\rm II}}  \\
  a^{\mathrm{II}}_{{\rm I},{\rm II},{\rm II};{\rm I},{\rm II}} \\
  a^{\mathrm{II}}_{{\rm I},{\rm II},{\rm II};{\rm II},{\rm II}}  \\
  a^{\mathrm{II}}_{{\rm II},{\rm II},{\rm II};{\rm I},{\rm II}}  \\  
  a^{\mathrm{II}}_{{\rm II},{\rm II},{\rm II};{\rm II},{\rm II}} \\
  \end{array} } \right]
  \equiv
  \left[ {\begin{array}{c}
  \frac{3(n^2-7n+8)}{(n-1)^1 (n-2)^5} \\
  \frac{n^5-15n^4+78n^3-165n^2+159n-62}{2(n-1)^2 (n-2)^5} \\
   \frac{3n^5-42n^4+211n^3-448n^2+436n-168}{4(n-1)^2 (n-2)^5} \\
  \frac{n^7-18n^6+127n^5-420n^4+574n^3-40n^2-608n+416}{8(n-1)^2 (n-2)^5} \\
 \frac{-n^5+19n^4-118n^3+296n^2-336n+148}{2(n-1)^2 (n-2)^5} \\
   \frac{-2n^5+41n^4-260n^3+659n^2-750n+328}{4(n-1)^2 (n-2)^5} \\
    \frac{3n^5-72n^4+531n^3-1494n^2+1848n-864}{8(n-1)^2 (n-2)^5} \\
    \frac{3n^6-39n^5+151n^4-45n^3-726n^2+1344n-736}{8(n-1)^2 (n-2)^5} \\
  \frac{n^7-14n^6+81n^5-352n^4+1412n^3-3384n^2+3984n-1824}{16(n-1)^2 (n-2)^5} \\
 \frac{3n^5-17n^4-25n^3+243n^2-420n+232}{2(n-1)^2 (n-2)^5} \\
  \frac{3n^6-24n^5+147n^4-1006n^3+3136n^2-4240n+2112}{16(n-1)^2 (n-2)^5} \\
   \frac{3n^8-47n^7+315n^6-1229n^5+3110n^4-4088n^3+336n^2+4928n-3648}{32(n-1)^2(n-2)^5} \\
    \end{array} } \right]\, .
\ee

\section{Nonperturbative RG}
\label{NRG}
We study the replicon field theory from the nonperturbative RG approach proposed by Wetterich~\cite{Wetterich93}. This scheme uses the Legendre transform of the Polchinski equation~\cite{Polchinski84}, casting the exact RG equations in a way that naturally leads to various approximation schemes. As such, it has had success with the Lifshitz critical point~\cite{Bervillier04}, random-field spin models~\cite{TT04,TT06,TT11}, fully-developed turbulent flows~\cite{CDW16}, and others~\cite{BTW02,Delamotte12}.

More specifically, within the nonperturbative RG scheme~\cite{BTW02,Delamotte12}, the microscopic action $S_{\Lambda}\le[\phi\ri]$ is supplemented by a cutoff term
\be
\Delta S\le[\phi\ri]=\frac{1}{2}\int \frac{\mathrm{d}\mathbf{q}}{(2\pi)^d} R_{\mu}\le(\mathbf{q}^2\ri)\sum_{i=1}^{\frac{n(n-3)}{2}}\phi_{i}\le(\mathbf{q}\ri)\phi_{i}\le(-\mathbf{q}\ri)\, ,
\ee
where the scale-dependent cutoff function, $R_{\mu}\le(\mathbf{q}^2\ri)$, suppresses \textit{low}-momentum fluctuations, \textit{i.e.} with $|\mathbf{q}|\lesssim \mu$ [\textit{cf.}~Eq.~(\ref{Litim})].
The resulting one-particle-irreducible effective action, $\Gamma_{\mu}\le[\phi\ri]$, then obeys the Wetterich equation~\cite{Wetterich93}
\be\label{Wet}
\mu\frac{\partial}{\partial\mu}\Gamma_{\mu}\le[\phi\ri]=\frac{1}{2}\int \frac{\mathrm{d}\mathbf{q}}{(2\pi)^d} \le\{\mu\frac{\partial}{\partial\mu}R_{\mu}\le(\mathbf{q}^2\ri)\ri\}\sum_{i=1}^{\frac{n(n-3)}{2}}\le\{\le(\Gamma^{(2)}_{\mu}\le[\phi\ri]+R_{\mu}\mathbbm{1}\ri)^{-1}\ri\}_{i,i}\le(\mathbf{q},\mathbf{q}\ri)\, ,
\ee
where
\be
\le(\Gamma^{(2)}_{\mu}\le[\phi\ri]\ri)_{i,j}\le(\mathbf{q},\mathbf{q}'\ri)\equiv\frac{\delta^2 \Gamma_{\mu}\le[\phi\ri]}{\delta\phi_{i}\le(-\mathbf{q}\ri)\delta\phi_{j}\le(\mathbf{q}'\ri)}\, 
\ee
and
\be
\le(R_{\mu}\mathbbm{1}\ri)_{i,j}\le(\mathbf{q},\mathbf{q}'\ri)\equiv R_{\mu}\le(\mathbf{q}^2\ri) \delta_{ij} \le(2\pi\ri)^d \delta^{(d)}\le(\mathbf{q}-\mathbf{q}'\ri)\, .
\ee
Although the Wetterich equation is exact, it is intractable in practice.
As mentioned above, it nonetheless provides a natural starting point for devising various approximation schemes.
We here adopt the most commonly employed scheme, the pseudo-local potential approximation, which implements the derivative expansion on the one-particle-irreducible effective action.
In order to make the analysis tractable in presence of complex index structures, we further truncate the potential-energy term. 
We find that the strict truncation to cubic order produces a behavior qualitatively similar to one-loop perturbative calculations without stable fixed points for $d<\du$, while the inclusion of quartic terms as independent couplings results in a plethora of spurious, unphysical, fixed points, as was also observed in simpler models~\cite{MOP88}.
In order to correctly treat higher-order contributions, we thus follow the systematic approach of Ref.~\cite{PW95}, which reproduces two-loop results for $d=\du$ when perturbatively expanded in couplings, while being similarly robust both at weak and strong couplings; see also Refs.~\cite{MOP88,Morris94} for different schemes.

Note that to fully imitate the approach of Ref.~\cite{PW95} and, in particular, to successfully reproduce the two-loop results in the weak-coupling limit we need to expand terms around the vacuum expectation value of a generic RSB phase and include more derivative terms. Because properly treating Nambu-Goldstone soft modes around a RSB phase remains an open problem, what follows is a simplified scheme. We nonetheless checked that the results are qualitatively robust against various changes of the scheme: (i) excluding the quintic term, $\frac{1}{5!}\sum_{i_1,\ldots,i_5=1}^{\frac{n(n-3)}{2}}\tilde{V}_{(5)\ast}^{i_1i_2i_3i_4i_5}\phi_{i_1}\phi_{i_2}\phi_{i_3}\phi_{i_4}\phi_{i_5}$; (ii) including the cubic term with two derivatives, $\frac{1}{2}\sum_{i_1,j_1,j_2=1}^{\frac{n(n-3)}{2}}\tilde{D}_{(3)\ast}^{i_1|j_1j_2}\phi_{i_1}\le(\nabla\phi_{j_1}\ri)\le(\nabla\phi_{j_2}\ri)$; (iii) including both cubic and quartic terms with two derivatives, the latter having the form of $\frac{1}{4}\sum_{i_1,i_2,j_1,j_2=1}^{\frac{n(n-3)}{2}}\tilde{D}_{(4)\ast}^{i_1i_2|j_1j_2}\phi_{i_1}\phi_{i_2}\le(\nabla\phi_{j_1}\ri)\le(\nabla\phi_{j_2}\ri)$; (iv) including all cubic, quartic, and quintic terms with two derivatives, the last having the form of $\frac{1}{12}\sum_{i_1,i_2,i_3,j_1,j_2=1}^{\frac{n(n-3)}{2}}\tilde{D}_{(5)\ast}^{i_1i_2,i_3|j_1j_2}\phi_{i_1}\phi_{i_2}\phi_{i_3}\le(\nabla\phi_{j_1}\ri)\le(\nabla\phi_{j_2}\ri)$; and (v) changing the sharp cutoff function, Eq.~(\ref{Litim}), to a smooth $R_{\mu}\le(\mathbf{q}^2\ri)=\frac{Z_{\mu} \mathbf{q}^2}{\mathrm{exp}\le(\mathbf{q}^2/\mu^2\ri)-1}$. Of these, only (iv) qualitatively changed the results, but this interference and the quantitative disagreement with other approaches mentioned in the main text would be likely cured if effects of the vacuum expectation value were properly included.

Within this approach, the effective action contains two parts, $\Gamma_{\mu}\le[\phi\ri]=\Gamma_{\mu}^{\mathrm{primary}}\le[\phi\ri]+\tilde{\Gamma}_{\ast}\le[\phi\ri]$.
The first is the primary action
\be
\Gamma_{\mu}^{\mathrm{primary}}\le[\phi\ri]=\int \mathrm{d}\mathbf{x}\le\{\frac{Z_{\mu}}{2}\sum_{i=1}^{\frac{n(n-3)}{2}}\le(\nabla\phi_i\ri)^2+\frac{\tilde{r}_{\mu}}{2}\sum_{i=1}^{\frac{n(n-3)}{2}}\phi_i^2-\frac{1}{3!}\sum_{i_1,i_2,i_3=1}^{\frac{n(n-3)}{2}}\le(\sum_{\T\in\le\{\mathrm{I},\mathrm{II}\ri\}}\tilde{g}_{\mu}^{\T}T_{\T}^{i_1i_2i_3}\ri)\phi_{i_1}\phi_{i_2}\phi_{i_3}\ri\}\, ,
\ee
governed by independent couplings, $\le\{Z_{\mu},\tilde{r}_{\mu},\tilde{g}_{\mu}^{\T}\ri\}$.
The second is the one-loop improved action
\be\label{slave}
\tilde{\Gamma}_{\ast}\le[\phi\ri]=\frac{1}{2}\int \frac{\mathrm{d}\mathbf{q}}{(2\pi)^d}\sum_{i=1}^{\frac{n(n-3)}{2}}\le\{\mathrm{ln}\le(\Gamma^{\mathrm{primary}(2)}_{\mu}\le[\phi\ri]+R_{\mu}\mathbbm{1}\ri)\ri\}_{i,i}\le(\mathbf{q},\mathbf{q}\ri)\, ,
\ee
from which we discard terms that are already contained in the primary action.
The secondary action can then be written as
\bea
\tilde{\Gamma}_{\ast}\le[\phi\ri]&=&\int \mathrm{d}\mathbf{x}\Bigg\{\frac{1}{4!}\sum_{i_1,i_2,i_3,i_4=1}^{\frac{n(n-3)}{2}}\tilde{V}_{(4)\ast}^{i_1i_2i_3i_4}\phi_{i_1}\phi_{i_2}\phi_{i_3}\phi_{i_4}-\frac{1}{5!}\sum_{i_1,\ldots,i_5=1}^{\frac{n(n-3)}{2}}\tilde{V}_{(5)\ast}^{i_1i_2i_3i_4i_5}\phi_{i_1}\phi_{i_2}\phi_{i_3}\phi_{i_4}\phi_{i_5}\Bigg\}\, .
\eea
Here, terms beyond the quintic order do not affect the renormalization group equations for independent couplings and are thus suppressed.

In order to express secondary couplings as functions of independent couplings, we first expand the logarithm in the prescription of Eq.~(\ref{slave}).
At $\ell$-th order in $\phi$, the one-loop improved action is given by
\bea
&&\frac{(-1)}{2\ell}\sum_{\T_1,\ldots,\T_{\ell}\in\le\{\mathrm{I},\mathrm{II}\ri\}}\sum_{i_1,\ldots,i_{\ell}=1}^{\frac{n(n-3)}{2}}\tilde{g}_{\mu}^{\T_1}\tilde{g}_{\mu}^{\T_2}\cdots\tilde{g}_{\mu}^{\T_{\ell}}\omega_{\T_1,\T_2,\ldots,\T_{\ell}}^{i_1i_2\ldots i_{\ell}}\\
&&\times\int \frac{\mathrm{d}\mathbf{q}_1}{(2\pi)^d}\frac{\mathrm{d}\mathbf{q}_2}{(2\pi)^d}\cdots \frac{\mathrm{d}\mathbf{q}_{\ell}}{(2\pi)^d}A_0\le(\mathbf{q}_1^2\ri)A_0\le(\mathbf{q}_2^2\ri)\cdots A_0\le(\mathbf{q}_{\ell}^2\ri)\phi_{i_1}\le(\mathbf{q}_{\ell}-\mathbf{q}_1\ri)\phi_{i_2}\le(\mathbf{q}_1-\mathbf{q}_2\ri)\cdots \phi_{i_{\ell}}\le(\mathbf{q}_{\ell-1}-\mathbf{q}_{\ell}\ri)\, ,\nonumber
\eea
with
\be
A_0\le(\mathbf{q}^2\ri)\equiv\frac{1}{Z_{\mu} \mathbf{q}^2+R_{\mu}\le(\mathbf{q}^2\ri)+\tilde{r}_{\mu}}\, 
\ee
and
\be
\omega_{\T_1,\T_2,\ldots,\T_{\ell}}^{i_1i_2\ldots i_{\ell}}\equiv\sum_{i_{\ell+1},\ldots,i_{2\ell}=1}^{\frac{n(n-3)}{2}}T_{\T_1}^{i_1i_{2\ell}i_{\ell+1}}T_{\T_2}^{i_2i_{\ell+1}i_{\ell+2}}\cdots T_{\T_{\ell}}^{i_{\ell}i_{2\ell-1}i_{2\ell}}\, .
\ee
Plugging in homogeneous field configurations then yields dimensionless secondary couplings
\bea
V_{(\ell)\ast}^{i_1i_2\ldots i_{\ell}}&\equiv&Z_{\mu}^{-\frac{\ell}{2}}\mu^{\frac{d(\ell-2)}{2}-\ell}\le(\frac{K_d}{c_d}\ri)^{\frac{(\ell-2)}{2}}\tilde{V}_{(\ell)\ast}^{i_1i_2\ldots i_{\ell}}\, ,
\eea
which we express in terms of the dimensionless independent couplings
\bea
r&\equiv& Z_{\mu}^{-1}\mu^{-2}\tilde{r}_{\mu}\, ,\\
g^{\T}&\equiv& Z_{\mu}^{-\frac{3}{2}}\mu^{\frac{d-6}{2}}\sqrt{\frac{K_d}{c_d}}\tilde{g}_{\mu}^{\T}\, .
\eea
Here,
\be\label{sphere}
K_{d}\equiv\frac{\mathrm{vol}(S^{d-1})}{(2\pi)^d}=\frac{1}{2^{d-1}\pi^{\frac{d}{2}}\Gamma\le(\frac{d}{2}\ri)}\, .
\ee
We shall later set the normalization constant, $c_d$, such that the final renormalization group equations agree with the perturbative equations when expanded to one-loop order.
Letting $(i_1i_2\ldots i_{\ell})$ denote the symmetric average over $\ell !$ permutations of indices, we obtain
\bea
V_{(4)\ast}^{i_1i_2i_3i_4}&=&-3s_3(r)\sum_{\T_1,\ldots,\T_4\in\le\{\mathrm{I},\mathrm{II}\ri\}}\omega^{(i_1i_2i_3i_4)}_{\T_1,\T_2,\T_3,\T_4}g^{\T_1}g^{\T_2}g^{\T_3}g^{\T_4}\, ,\\
V_{(5)\ast}^{i_1i_2i_3i_4i_5}&=&12s_4(r)\sum_{\T_1,\ldots,\T_5\in\le\{\mathrm{I},\mathrm{II}\ri\}}\omega^{(i_1i_2i_3i_4i_5)}_{\T_1,\T_2,\T_3,\T_4,\T_5}g^{\T_1}g^{\T_2}g^{\T_3}g^{\T_4}g^{\T_5}\, ,
\eea
where we have introduced the functions
\be
s_{\ell}\le(r\ri)\equiv \frac{ c_d}{2}\int_{0}^{\infty} \mathrm{d}y y^{\frac{d}{2}-1}\frac{1}{\le\{y+b(y)+r\ri\}^{\ell+1}}\, ,
\ee
with
\be
b(y)\equiv\frac{1}{Z_{\mu} \mu^2}R_{\mu}(\mathbf{q}^2=\mu^2 y)\, .
\ee

In order to evaluate the right-hand side of the Wetterich equation~(\ref{Wet}), we need to invert the matrix
\be
\le\{\Gamma^{(2)}_{\mu}\le[\phi\ri]+R_{\mu}\mathbbm{1}\ri\}_{i,j}\le(\mathbf{q},\mathbf{q}'\ri)
\ee
to cubic order in $\phi$ and evaluate diagonal elements.
Along with the identity $\sum_{i=1}^{\frac{n(n-3)}{2}}T_{\T}^{iij}=0$, the following combinatorial relations prove useful in performing the algebra:
\bea
\sum_{i_3=1}^{\frac{n(n-3)}{2}}\sum_{\T_1,\T_2,\T_3,\T_4\in\le\{\mathrm{I},\mathrm{II}\ri\}}g^{\T_1}g^{\T_2}g^{\T_3}g^{\T_4}\omega^{(i_1i_2i_3i_3)}_{\T_1,\T_2,\T_3,\T_4}&=&\delta^{i_1i_2}\le(\frac{2}{3}I_{2}^2+\frac{1}{3}I_{4}\ri)\, ,\\
\sum_{i_4,i_5=1}^{\frac{n(n-3)}{2}}\sum_{\T_1,\T_2,\T_3,\T_4,\T_5\in\le\{\mathrm{I},\mathrm{II}\ri\}}g^{\T_1}g^{\T_2}g^{\T_3}g^{\T_4}g^{\T_5}\omega^{(i_1i_2i_4i_5)}_{\T_1,\T_2,\T_3,\T_4}T_{\T_5}^{i_3i_4i_5}&=&\sum_{\T\in\le\{\mathrm{I},\mathrm{II}\ri\}}T_{\T}^{i_1i_2i_3}\le(\frac{1}{3}I_{5,A}^{\T}+\frac{2}{3}I_{5,B}^{\T}\ri)\, ,\\
\sum_{i_4=1}^{\frac{n(n-3)}{2}}\sum_{\T_1,\T_2,\T_3,\T_4,\T_5\in\le\{\mathrm{I},\mathrm{II}\ri\}}g^{\T_1}g^{\T_2}g^{\T_3}g^{\T_4}g^{\T_5}\omega^{(i_1i_2i_3i_4i_4)}_{\T_1,\T_2,\T_3,\T_4,\T_5}&=&\sum_{\T\in\le\{\mathrm{I},\mathrm{II}\ri\}}T_{\T}^{i_1i_2i_3}\le(\frac{1}{2}I_{2}I_{3}^{\T}+\frac{1}{2}I_{5,B}^{\T}\ri)\, ,
\eea
all of which can be derived by contracting indices of appropriate tensor products. We further define threshold functions
\bea
f_{\ell}(r)&\equiv&\frac{\ell c_d}{2}\int_{0}^{\infty} \mathrm{d}y y^{\frac{d}{2}-1}\frac{c(y)}{\le\{y+b(y)+r\ri\}^{\ell+1}}\, ,\\
m_{1}(r)&\equiv& c_d\int_{0}^{\infty} \mathrm{d}y y^{\frac{d}{2}-1}c(y)\le[\frac{1+b'(y)+\frac{2y}{d}b''(y)}{\le\{y+b(y)+r\ri\}^{4}}-\frac{4y}{d}\frac{\le\{1+b'(y)\ri\}^2}{\le\{y+b(y)+r\ri\}^{5}}\ri]\, ,
\eea
with
\be
c(y)\equiv\frac{1}{Z_{\mu} \mu^2}\le(\mu\frac{\partial R_{\mu}}{\partial \mu}\ri)(\mathbf{q}^2=\mu^2 y)\, .
\ee
Noting that the anomalous exponent is given by
\be
\eta= -\mu\frac{\partial \mathrm{log}\le(Z_{\mu}\ri)}{\partial\mu}\, ,
\ee
the resulting nonperturbative RG equations can be written as
\bea
\eta&=&\frac{1}{2}{m}_{1}\le(r\ri)I_2(g)\, ,\\
\beta_{r}\equiv\mu\frac{\partial r}{\partial \mu}&=&(-2+\eta)r+\frac{1}{2}f_{2}(r)I_2(g)+\le\{f_{1}(r)s_{3}(r)\ri\}I_{2}^2(g)+\le\{\frac{1}{2}f_{1}(r)s_{3}(r)\ri\}I_{4}(g)\, ,\\
\beta^{\T}\equiv\mu\frac{\partial g^{\T}}{\partial \mu}&=&\le(\frac{d-6+3\eta}{2}\ri)g^{\T}-f_{3}(r)I^{\T}_{3}(g)\\
&&-\le\{3f_{1}(r)s_{4}(r)\ri\}I_{2}(g)I^{\T}_{3}(g)-\le\{\frac{3}{2}f_{2}(r)s_{3}(r)\ri\}I^{\T}_{5,A}(g)-\le\{3f_{2}(r)s_{3}(r)+3f_{1}(r)s_{4}(r)\ri\}I^{\T}_{5,B}(g)\, .\nonumber
\eea

In order to numerically study these equations, we selected the cutoff function
\be\label{Litim}
R_{\mu}\le(\mathbf{q}^2\ri)=Z_{\mu}\le(\mu^2-\mathbf{q}^2\ri)\theta\le(\mu^2-\mathbf{q}^2\ri)\, 
\ee
and the normalization constant
\be
c_d=\frac{d}{6}\, ,
\ee
which give
\bea
f_{\ell}(r)&=&\frac{\ell}{3\le(1+r\ri)^{\ell+1}}\le(1-\frac{\eta}{d+2}\ri)\, ,\\
m_{1}(r)&=&\frac{1}{3\le(1+r\ri)^{4}}\, ,\\
s_{\ell}\le(r\ri)&=&\frac{1}{6(1+r)^{\ell+1}}+\frac{d}{6(2\ell+2-d)}\ _2F_{1}\le(\ell+1,\ell+1-\frac{d}{2};\ell+2-\frac{d}{2};-r\ri)\, ,
\eea
where we used
\be
\int_0^{\infty}\mathrm{d}y \theta(1-y)\delta(1-y)=\frac{1}{2}\, .
\ee
Note that in general there is subtlety in dealing with products of step and delta functions~\cite{Morris94}, but within our approximation, such subtleties do not arise.

Within this nonperturbative approach, unlike the perturbative dimensional regularization scheme, there is no clean way to focus on the critical surface from the onset. The analysis must instead include the quadratic coupling, $r(\mu)$, which essentially corresponds to the relevant deformation of the system away from the critical point. The RG flow is thus governed by three $\beta$-functions, $\le\{\beta_r, \beta^{\mathrm{I}},\beta^{\mathrm{II}}\ri\}$, and the critical surface is defined by the global condition that a flow starting at $\le\{r,g^{\mathrm{I}},g^{\mathrm{II}}\ri\}$ is attracted to the critical fixed point (or the critical limit cycle). In other words, a codimension-one hypersurface, $r_{\rm c}\le(g^{\mathrm{I}},g^{\mathrm{II}}\ri)$, is identified over the range of $(g^{\mathrm{I}},g^{\mathrm{II}})$ that can be made critical by tuning $r$, as long as the fixed point (or the cycle) remains critical with a single relevant deformation. The stability exponents of the fixed point are given by the right eigenvalues of a $3\times3$ matrix
\be
\left[ {\begin{array}{ccc}
 \frac{\partial\beta_r}{\partial r} & \frac{\partial\beta_r}{\partial g^{\mathrm{I}}} &  \frac{\partial\beta_r}{\partial g^{\mathrm{II}}} \\
  \frac{\partial\beta^{\mathrm{I}}}{\partial r} & \frac{\partial\beta^{\mathrm{I}}}{\partial g^{\mathrm{I}}} &  \frac{\partial\beta^{\mathrm{I}}}{\partial g^{\mathrm{II}}} \\
  \frac{\partial\beta^{\mathrm{II}}}{\partial r} &  \frac{\partial\beta^{\mathrm{II}}}{\partial g^{\mathrm{I}}} & \frac{\partial\beta^{\mathrm{II}}}{\partial g^{\mathrm{II}}} \\
  \end{array} } \right]\Bigg|_{(r,g^{\mathrm{I}},g^{\mathrm{II}})=(r_{\star},g_{\star}^{\mathrm{I}},g_{\star}^{\mathrm{II}})}\, ,
\ee
with the lowest value, $\lambda_0$, yielding the critical exponent, $\nu=-1/\lambda_0$, while $\lambda_{1}$ and $\lambda_2$ again control subleading corrections near the critical point. We find that the stability exponents behave qualitatively similar to those of the minimal two-loop RG, but the numerical analysis of the limit cycle becomes arduous due to the need for manually tuning out one relevant deformation.

\end{widetext}

\bibliography{dATGm}

\begin{thebibliography}{79}
\bibitem [{foo()}]{footnoteRK}%
  \BibitemOpen
  \href@noop {} {}\bibinfo {note} {We used the fourth-order Runge-Kutta
  integration with timestep
  $\mathrm{d}t\equiv\mathrm{d}\mu/\mu=10^{-5}$.}\BibitemShut {Stop}%
\end{thebibliography}

\end{document}